\documentclass[prx,twocolumn,floatfix,superscriptaddress,amsmath,aps,longbibliography]{revtex4-1}
\usepackage{dcolumn,graphicx,color,booktabs,microtype,afterpage}
\usepackage[colorlinks,plainpages=false,linkcolor=blue,urlcolor=blue,citecolor=blue,pdfpagemode=UseNone,pdfstartview=FitBH]{hyperref}
\usepackage{graphicx}
\usepackage{color}
\usepackage{latexsym}

\begin{document}
\title{Field-induced magnetic transition and spin fluctuation  in quantum spin liquid candidate CsYbSe$_2$}
\author{Jie Xing}
\affiliation{Materials Science and Technology Division, Oak Ridge National Laboratory, Oak Ridge, Tennessee 37831, USA}
\author{Liurukara D. Sanjeewa}
\affiliation{Materials Science and Technology Division, Oak Ridge National Laboratory, Oak Ridge, Tennessee 37831, USA}
\author{Jungsoo Kim}
\affiliation{Department of Physics, University of Florida, Gainesville, Florida 32611, USA}
\author{G. R. Stewart}
\affiliation{Department of Physics, University of Florida, Gainesville, Florida 32611, USA}
\author{Andrey Podlesnyak}
\affiliation{Neutron Scattering Division, Oak Ridge National Laboratory, Oak Ridge, TN 37831}
\author{Athena S. Sefat}
\affiliation{Materials Science and Technology Division, Oak Ridge National Laboratory, Oak Ridge, Tennessee 37831, USA}

\date{\today}

\begin{abstract}
Two-dimensional triangular-lattice materials with spin-1/2 are perfect platforms for investigating quantum frustrated physics with spin fluctuations.
Here we report the structure, magnetization, heat capacity and inelastic neutron scattering (INS) results on cesium ytterbium diselenide, CsYbSe$_2$.
There is no long-range magnetic order down to 0.4~K at zero field.
The temperature dependent magnetization, $M(T)$, reveals an easy-plane magnetic anisotropy.
A maximum is found in $M(T)$ around \emph{T}$\sim$1.5~K when magnetic field $H$ is applied in the $ab$ plane, indicating the short-range interaction.
The low-temperature isothermal magnetization $M(H)$ shows a one-third plateau of the estimated saturation moment, that is characteristic of a two-dimensional frustrated triangular lattice.
Heat capacity shows field-induced long-range magnetic order for both $H||c$ and $H||ab$ directions.
The broad peak in heat capacity and highly damped INS magnetic excitation at $T$=2~K suggests strong spin fluctuations.
The dispersive in-plane INS, centered at the (1/3~1/3~0) point, and the absence of dispersion along $c$ direction suggests 120$^{\circ}$ non-collinear 2D-like spin correlations.
All these results indicate that the two-dimensional frustrated material CsYbSe$_2$ can be in proximity
to the triangular-lattice quantum spin liquid.
We propose an experimental low-temperature $H$-$T$ phase diagram for CsYbSe$_2$.
\end{abstract}

\maketitle

\section{Introduction}
Frustrated magnetism is a challenge and intriguing field in the condensed matter physics due to multiple unconventional phenomena having ground state degeneracy~\cite{sadoc_mosseri_1999,Moessner}.
One important topic is the quantum spin liquid (QSL) state, where highly entangled spins prevent to break any symmetry even at zero temperature~\cite{Anderson,Mila2000,Balents}.
The spin interactions are restricted by the low-dimensional structure, which could enhance the spin fluctuations~\cite{Moessner}.
Until now most QSL candidates are proposed in the low-spin $S$=1/2 frustrated systems, such as \emph{A}$_2$IrO$_3$ (\emph{A}=Na, Li, Cu), H$_3$LiIr$_2$O$_6$, $\kappa$-(BEDT-TTF)$_2$Cu$_2$(CN)$_3$, EtMe$_3$Sb[Pd(dmit)$_2$]$_2$, and RuCl$_3$~\cite{singh2010antiferromagnetic, singh2012relevance, chaloupka2010kitaev, mazin20122, ye2012direct, takayama2015hyperhoneycomb, chun2015direct, kitagawa2018spin, abramchuk2017cu2iro3, itou2008quantum, banerjee2016proximate, banerjee2017neutron, cao2016low, ran2017spin, leahy2017anomalous}.

The rare-earth materials attract an attention in search for QSL candidates, as an alternative to Cu-based $S$=1/2 systems.
The rare-earth ions with the odd number of 4\emph{f} electrons could be treated as Kramers doublets with an effective spin $S_{\mathrm{eff}}$=1/2.
Especially a number of Yb-based quantum magnets have attracted considerable interest, like the realization of a quantum spin $S$=1/2 chain in YbAlO$_3$~\cite{Wu1,Wu2,agrapidis} and a quantum dimer magnet Yb$_2$Si$_2$O$_7$~\cite{Hester}.
Another Yb-base compound, YbMgGaO$_4$, was proposed as QSL candidate with frustrated Yb$^{3+}$ triangular lattice.
The heat capacity, magnetization, thermal conductivity, neutron scattering and muon spin relaxation find no transition in this compound and suggest a possible gapless QSL ground state~\cite{li2015gapless,li2015rare,li2016muon,shen2016evidence,paddison2017continuous,zhang2018hierarchy,li2017crystalline,steinhardt2019field}.
However, the intrinsic Mg/Ga disorder exists between the frustrated layer and could induce the system to other state~\cite{li2015gapless,paddison2017continuous,zhang2018hierarchy,shen2018fractionalized,zhu2017disorder,zhu2018topography,kimchi2018valence}.

Another classic 112 system \emph{A}\emph{R}\emph{Q}$_2$(\emph{A}=Alkali metal, \emph{R}=Rare-earth elements, \emph{Q}= O, S, Se) was proposed as QSL candidate\cite{liu2018rare}.
This 112 system is unique with the perfect rare-earth triangular layers, which are separated by the alkali metal ions.
The distance of the nearest neighbor (NN) rare-earth ions and interlayer distance could be tuned by replacing the different \emph{A}$^+$ and \emph{Q}$^{2-}$ ions.
There is no structural or magnetic transition in NaYb\emph{Ch}$_2$ (\emph{Ch}=O,S,Se) polycrystalline down to 50~mK~\cite{liu2018rare}.
The quantum chemistry calculation suggests that the Yb$^{3+}$ 4\emph{f}$^{13}$ configuration should show large g$_{ab}$ factors within the frustrated magnetic layers~\cite{zangeneh2019single}.
Electron spin resonance and magnetization reveal $S_{\mathrm{eff}}$=1/2 ground state at low temperature in the NaYbS$_2$ \cite{baenitz2018naybs}. Until now, the investigation of 112 system mostly focuses on the NaYb$Q_2$ compounds with the rhombohedral lattice (space group $R\bar{3}m$).

Magnetic fields perturb the quantum disordered ground state and induce the long-range order up-up-down spin state in NaYbO$_2$~\cite{ding2019gapless,ranjith2019field,bordelon2019field}.
This field induced quantum phase transition, evoking alternative ground states, is an intriguing phenomenon in the frustrated triangular system.
Two-dimensional triangular-lattice Heisenberg antiferromagnet could present several ground states under different fields due to degeneracy, such as 120$^\circ$ spin state, or collinear up-up-down state~\cite{huse1988simple,chubokov1991quantum}.
In the easy-plane scenario, the up-up-down state could induce a plateau at one-third of the saturation magnetization due to the spin fluctuations~\cite{yamamoto2014quantum}, as it was found in the Cs$_2$CuBr$_4$, CuFeO$_2$, Ba$_3$CoNb$_2$O$_9$ and Ba$_3$CoSb$_2$O$_9$~\cite{ono2003magnetization,terada2006field,shirata2012experimental,lee2014series,Ito2017}. The phase diagrams in the quantum scenario are not well elucidated which inspires us to investigate more triangular-lattice compounds with different environments.

Here, we study another triangular lattice material CsYbSe$_2$.
We present the detailed measurements of the single crystal magnetization, heat capacity and inelastic neutron scattering (INS).
There is no long-range order in the magnetization and heat capacity down to $T$=0.4~K at zero field, while large anisotropy is found between $ab$ plane and $c$ axis.
The magnetic field induces a one-third plateau in the isothermal magnetization below \emph{H}$<$7~T.
Zero field heat capacity suggests the QSL state in the triangular Heisenberg lattice.
The low-temperature heat capacity with magnetic fields confirms the quantum-induced magnetic ordering in intermediate fields.
The excitation spectra obtained from INS unambiguously demonstrate a quasi-2D nature of frustrated Yb$^{3+}$ with $S_\mathrm{eff}$=1/2 in triangular-lattice.

\section{Materials and Methods}

\begin{figure}[tbh]
\includegraphics[width=1\linewidth]{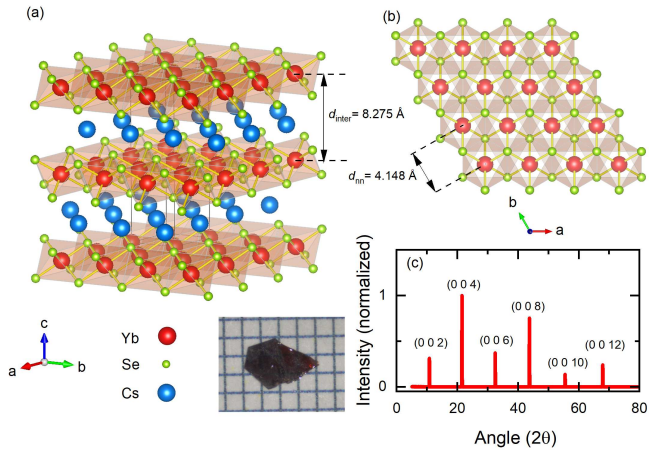}
\caption {(a) The structure of CsYbSe$_2$. The distance between the nearest two dimensional Yb triangular layers is 8.275 {\AA}. (b) The perfect frustrated Yb triangular layer. The distance of the nearest neighbor Yb ions is 4.148 {\AA}. Each Yb ion is connected to six nearest Yb ions via six Yb-Se-Yb bonds. (c) The x-ray diffraction pattern obtained on the surface of a single crystal.}
\label{structure}
\end{figure}

Millimeter-sized hexagonal shape CsYbSe$_2$ single crystals were synthesized by salt flux method following the procedure described in Ref.~\cite{xing2019}. The energy-dispersive x-ray spectroscopy (EDS) shows the molar ratio of Cs:Yb:Se is close to 1:1:2. Fig~\ref{structure}(a) demonstrates the structure and the optical microscope image of CsYbSe$_2$. X-ray diffraction (XRD) was collected on a PANalytical X'pert Pro diffractometer equipped with an incident beam monochromator (Cu K$\alpha_1$ radiation) at room temperature [Fig.~\ref{structure}(c)]. Sharp (00l) peaks suggest good crystalline quality in the single crystal.

Magnetic properties were measured using a Quantum Design (QD) Magnetic Properties Measurement System (MPMS3). The magnetization below 2~K was measured by MPMS3 iHe3 option. Temperature dependent heat capacity was measured in QD Physical Properties Measurement System (PPMS) using the relaxation technique.

Neutron scattering measurements of CsYbSe$_2$ were performed at the time-of-flight Cold Neutron Chopper Spectrometer (CNCS)~\cite{CNCS1,CNCS2}, at the Spallation Neutron Source at Oak Ridge National Laboratory.
Data were collected with 12 single crystal samples of total mass around 0.1~g, which were co-aligned to within 2$^{\circ}$ using a Multiwire x-ray Laue machine in the $(HHL)$ scattering plane.
The measurements were carried out using the rotating single crystal method at temperature of $T=$~2~K.
The data were collected using a fixed incident neutron energies of $E_i=3.32$~meV and $E_i=25.0$~meV resulting in a full-width at half-maximum energy resolution of 0.07~meV and 0.75~meV at the elastic position, respectively.
All time-of-flight data-sets were combined to produce a four-dimensional scattering-intensity function $I(\mathbf{Q},E_f)$, where $\mathbf{Q}$ is the momentum transfer and $E_f$ is the energy transfer.
For data reduction and analysis, we used the \textsc{Mantid}~\cite{Mantid} and \textsc{Horace}~\cite{Horace} software packages.

\section{Results and Discussion}

\begin{figure}[tbh]
\includegraphics[width=0.95\linewidth]{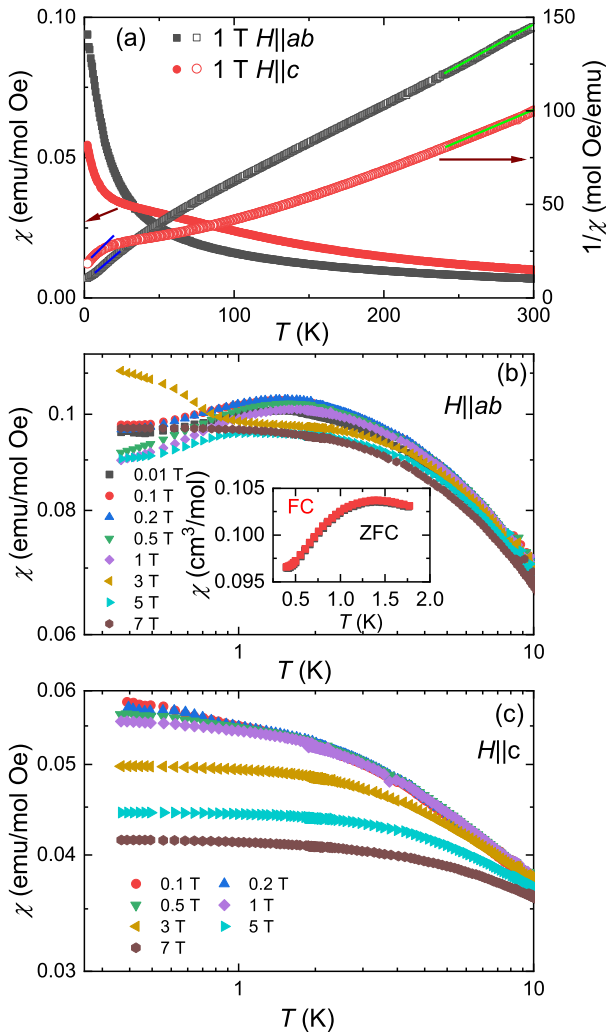}
\caption {(a) The temperature dependence of DC susceptibility for the CsYbSe$_2$ single crystal in \emph{ab} plane and along \emph{c} axis. Solid lines show the fit to the Curie-Weiss law, see the text and Table~\ref{Tab1}. (b,c) The temperature dependence of magnetic susceptibility in different magnetic fields $H||ab$ (b) $H||c$ (c). The discontinued points around 1.8~K are due to the different temperature control method of the measurement.
Inset of (b) shows the low temperature field cooled (FC) and zero field cooled (ZFC) susceptibility at $H$=0.2~T.}
\label{susc}
\end{figure}

Unlike NaYbO$_2$, which is crystallized in $R\bar{3}m$ space group, CsYbSe$_2$ adopts \emph{P6$_3$/mmc} space group due to different layer stacking sequence.
Because the large radius of Cs ions, two layers of Yb-Se-Yb are packed in CsYbSe$_2$ along $c$ axis while three layers of Yb-O-Yb are packed in NaYbO$_2$.
CsYbSe$_2$ has similar ideal triangular layers of Yb$^{3+}$ in the YbSe$_6$ octahedral environment, as shown in Fig.~\ref{structure}(b).
The distance between Yb-layers $d_{\mathrm{inter}}$ in CsYbSe$_2$ is 8.275 {\AA}, which is larger than the $d_{\mathrm{inter}}$ in NaYbO$_2$ and comparable to $d_{\mathrm{inter}}$ in YbMgGaO$_4$.
The distance between the nearest neighbour Yb$^{3+}$ $d_{\mathrm{NN}}$ is 4.148 {\AA}, which is slightly larger than $d_{\mathrm{NN}}$ in NaYbO$_2$ and NaYbS$_2$ due to larger size of Se ion.
The 2D ratio $d_{\mathrm{inter}}/d_{\mathrm{NN}}=2$ varies from 1.6 for NaYbO$_2$ to 2.47 for YbMgGaO$_4$.
This difference of the 2D ratio reflects on the different field-induced magnetic transition at low temperature in NaYbO$_2$ and YbMgGaO$_4$~\cite{ranjith2019field,bordelon2019field,steinhardt2019field}.
Based on two-dimensional CsYbSe$_2$ structure, it inspires us to investigate the details of magnetic properties using the single crystals.

\subsection{Magnetization}

The DC susceptibility measurements under $H$=1~T [Fig.~\ref{susc}(a)] show no evidence of a magnetic transition in the CsYbSe$_2$ down to 2~K.
We used Curie-Weiss law to fit the magnetization, in both field directions, for the temperatures above 250~K and around 10~K [Fig.~\ref{susc}(a)].
The effective magnetic moment $\mu_{\mathrm{eff}}$ was evaluated using the Curie constant from the fits.
The Curie-Weiss temperature $\theta_{CW}$ and $\mu_{\mathrm{eff}}$ are shown in Table~\ref{Tab1}.
At high temperatures, the effective moment is close to the moment of free Yb$^{3+}$ (4.54~$\mu_{\mathrm{B}}$).
As the temperature decreases, the effective moment also decreases and exhibits strong anisotropic behavior, similar to that found in NaYbS$_2$~\cite{baenitz2018naybs}.
In the anisotropic two dimensional Heisenberg system, the average magnetic susceptibility is described by $\chi_{avg}=(2\chi_{ab}+\chi_{c})/3$~\cite{li2015rare}.
The negative value $\theta_{avg}=-16.9$ K from Curie-Weiss fitting at low temperature indicates the antiferromagnetic interaction between Yb$^{3+}$ ions.
Interestingly, this value is slight lower than that $\theta=-10.3$ K found in similar Yb-triangular lattice materials NaYbO$_2$, although the $d_{\mathrm{NN}}$ in CsYbSe$_2$ is longer than $d_{\mathrm{NN}}$ in NaYbO$_2$.
\begin{table}[tbh]
\caption{The effective magnetic moment ($\mu_{\mathrm{eff}}$) and Curie-Weiss temperature ($\theta_{CW}$) obtained from the fit at high temperatures (HT) of $250-300$~K, and at low temperatures (LT) of $8-12$~K.}
\begin{ruledtabular}
\begin{tabular}{ccccc}
 & $\mu_{\mathrm{eff}}$[$\mu_{\mathrm{B}}$] & $\theta_{CW}$[K] & $\mu_{\mathrm{eff}}$[$\mu_{\mathrm{B}}$] & $\theta_{CW}$[K]  \\
 & HT & HT & LT & LT \\
\hline
$H\|$\emph{ab}  & 4.34  &  -42.3 & 3.48 &-13.2 \\
$H\|$\emph{c}   & 4.88  &  -31.6 &3.26  & -26.1 \\
\end{tabular}
\end{ruledtabular}
\label{Tab1}
\end{table}

The low-temperature part of the DC susceptibility is shown in Fig.~\ref{susc}(b,c).
There is no long-range order down to 0.42~K in both field directions below 1~T.
The larger $H\|$\emph{ab} magnetization indicates an easy-plane feature of CsYbSe$_2$, which is also found in other 112 compounds~\cite{baenitz2018naybs,jie2019naerse}.
A maximum is found in the temperature dependence of magnetization below 3~T for the $H\|$\emph{ab}.
Since the heat capacity does not exhibit a $\lambda$ anomaly in this temperature range, we suggest that the strong short-range interactions are developed.
Zero-field cooled (ZFC) and field cooled (FC) susceptibility overlap at low temperatures (see inset in Fig.~\ref{susc}(b)), excluding the possibility of spin glass state.
This behavior is different from that observed in NaYbS$_2$ single crystal, indicating a possible difference in the in-plane magnetism. It may be caused by the different distances between Yb$^{3+}$, spin-orbital coupling or different space groups.
When the magnetic field reaches 3~T, an upturn is found at $\sim$0.8~K, revealing the field-induced long-range magnetic order (LRO).
As the magnetic field reaches 7~T, the magnetization shows a flat feature, indicating that the system moves to another magnetic state.
Both LRO and maximum near 1.5~K were not found when magnetic field up to 7~T is applied along $c$ axis due to the easy-plane anisotropy.
The heat capacity measurements reveal the field-induced magnetic transition around 10~T, as we discuss in Section~\ref{heatcapacity}.
Although a similar maximum is not found for the $H\|$\emph{c}, the observed deviation of the magnetic susceptibility from Curie-Weiss law below 8~K suggests that the crossover to QSL is present.

\begin{figure}[tbh]
\includegraphics[width=0.95\linewidth]{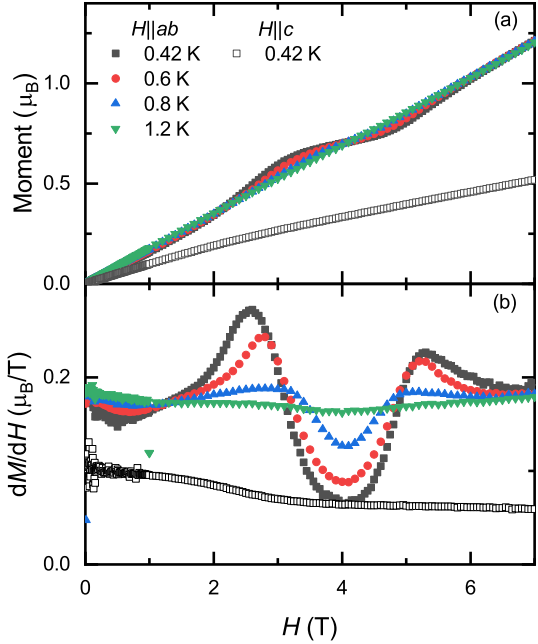}
\caption {Magnetization (a) and differential of magnetization (b) vs field for \emph{H}$\|$\emph{ab} and \emph{H}$\|$\emph{c} at different temperatures.}
\label{magn}
\end{figure}

The effect of an applied field on the magnetization at different temperatures is shown in Fig.~\ref{magn}(a).
The observed anisotropic ratio is larger than in NaYbS$_2$, which is consistent with the larger $d_{\mathrm{inter}}/d_{\mathrm{NN}}$ ratio in CsYbSe$_2$.
We found an apparent magnetic plateau below $T$=1.2~K for the magnetic moment $\sim$0.7 $\mu_{\mathrm{B}}$.
From the theoretical calculations, the 1/3 magnetic plateau is expected in the easy-plane XXZ model~\cite{yamamoto2014quantum}.
Using the Curie-Weiss fitting at low temperature, we expect the $g$ value $g_{ab}=$~4.0 by assuming $S_{\mathrm{eff}}$=1/2.
So, we estimate the saturation moment at low temperature $m_s = gJ$ to be 2~$\mu_{\mathrm{B}}$.
The observed magnetic plateau is indeed close to 1/3 of the estimated saturation moment.
There is no similar 1/3-plateau on the magnetization curve in field $H||c$ up to 7~T.
This is caused by the easy-plane anisotropy, which requires a higher magnetic field to build up a long-rang order along $c$ axis.
The differential of the magnetization shows two peaks for the field direction $H\|$\emph{ab} (Fig.~\ref{magn}(b)).
They become broader with temperature increasing but we still observe the anomalies at $T=1.2$~K at 3~T and 5~T, indicating strong spin fluctuation.
No peaks were found in differential of magnetization for the field direction $H||c$ up to $H=7$~T.

\subsection{Heat Capacity}\label{heatcapacity}

\begin{figure}[tb]
\includegraphics[width=0.95\linewidth]{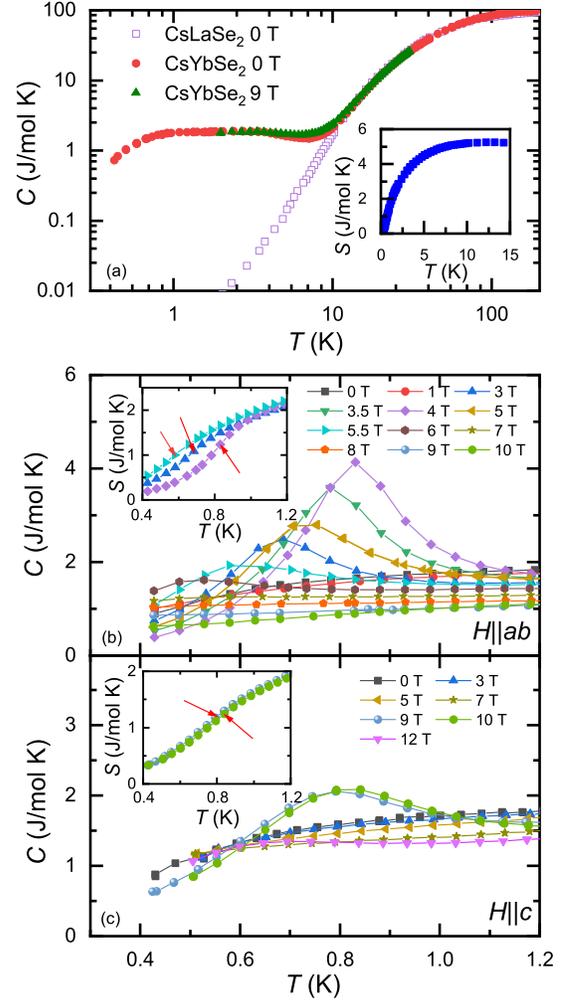}
\caption {(a) The temperature dependence of heat capacity of  CsYbSe$_2$ and CsLaSe$_2$. The 9~T magnetic field is applied along \emph{c} axis. The inset presents the magnetic entropy of  CsYbSe$_2$ below 15~K. (b,c) Low temperature specific heat of CsYbSe$_2$ under different magnetic fields for $H||ab$ (b) and $H||c$ (c) axis. The inset shows the magnetic entropy of CsYbSe$_2$ in different magnetic fields below 1.2~K. The red arrow indicates the positions of LRO at each field.}
\label{hc}
\end{figure}

Zero-field heat capacity for CsYbSe$_2$ is shown in Fig.~\ref{hc}(a).
The heat capacity goes to Dulong-Petit limit of phonon contribution around 100 J/mol K~\cite{kittel1976solid}.
A broad peak is found below 10~K, which is similar to that observed in the other Yb-112 materials implying the QSL states in the triangular Heisenberg lattices~\cite{bordelon2019field,wang1992specific,isoda2011specific}.
There is no $\lambda$ anomaly around 1.5~K, confirming short range correlations found in the magnetization.
The broad peak moves only slightly to high temperature in field H$\|$\emph{c} of 9~T (see Fig.~\ref{hc}(a)).
The shift is smaller than that in NaYbO$_2$ powder due to easy-plane feature or large distance between the nearest neighbour Yb$^{3+}$.

To estimate the magnetic entropy, we subtract the phonon background using a nonmagnetic CsLaSe$_2$, as shown in Fig.~\ref{hc}(a).
The integrated entropy is 5.4~J/mol K, which is 93\% of the Rln2 of spin-1/2 system.
This result is consistent with the $S_{\mathrm{eff}}$=1/2 doublet in CsYbSe$_2$.

To investigate the field-induced magnetic phase transition, we measured the heat capacity in field of two directions, $H||ab$ and $H||c$, at different temperatures, as shown in Fig.~\ref{hc}(b-c).
When the small magnetic field was applied along $ab$ plane and $c$ axis, the heat capacity curves overlap very well,  suggesting spin liquid behavior at the low magnetic fields.
The field-induced magnetic transition is observed at $H=3$~T when the magnetic field was applied along the $ab$ plane,that is consistent with the magnetization measurements.
The temperature and magnitude of the $\lambda$ anomaly increase with the increasing of the magnetic field up to 4~T.
The anomaly has the sharpest shape at the highest transition temperature $\sim$0.8~K and a long tail above 1~K, which indicates the large spin fluctuations.
The magnetic transition is suppressed to a low temperature in fields $H||ab$ above 4~T and disappeared at 7~T.
The heat capacity at high magnetic field ($>$7~T) shows field-dependent relation and clear difference from the low field ones ($<$3~T), indicating the different magnetic states in these two regions.

When increasing the $H||c$ magnetic field up to 9~T, a field-induced magnetic transition appears at 0.8~K that is similar to the case of  $H||ab$ at 4~T.
If we extend $M$($H$) to 9 T at $H||c$ and pick $M$($H$) at 3 T at $H||ab$, the estimated values of the magnetic moments at the magnetic transition are close in both directions. The magnetic field suppresses the transition to 0.6~K at 12~T and clear difference was found at 3~T and 12~T above 0.6~K.
This feature also indicates the different states induced by the high magnetic fields.
The anisotropic heat capacity under magnetic fields is consistent with the magnetization measurements, suggesting the two-dimensional magnetism in CsYbSe$_2$ with the easy-plane anisotropy.
The magnetic transition temperature in CsYbSe$_2$ is slightly lower than in the NaYbO$_2$. 
The integrated entropy in the vicinity of the LRO transition, as estimated in the inset of Fig.~\ref{hc}(b,c), is below 20\% of Rln2, revealing strong spin fluctuation in CsYbSe$_2$.

\begin{figure}[tb]
\includegraphics[width=1\linewidth]{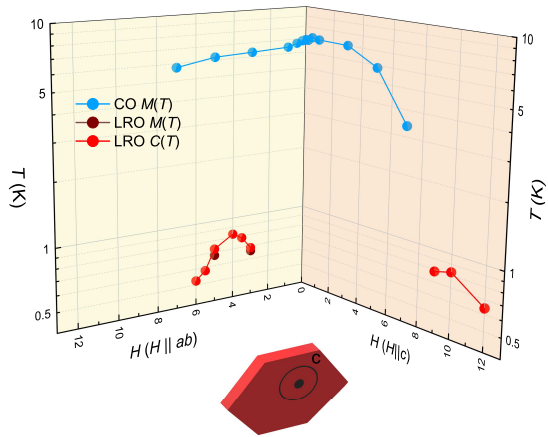}
\caption {Experimental $H$-$T$ phase diagram of CsYbSe$_2$ obtained from heat capacity and magnetic susceptibility results. The values of the crossover (CO) are extracted as the temperature where magnetic susceptibility deviates from Curie-Weiss relation.}
\label{phasediagram}
\end{figure}

Based on the magnetization and heat capacity results, in Fig.~\ref{phasediagram} we present a tentative $H$-$T$ phase diagram for CsYbSe$_2$. Large magnetic anisotropy is easily seen.
The values of the crossover (CO) are extracted as the temperature where magnetic susceptibility deviates from Curie-Weiss relation.
The LRO values obtained from heat capacity and magnetic susceptibility are consistent with each other.
Note that the dome feature for LRO is also found in other frustrated magnetic materials~\cite{bordelon2019field,quirion2015magnetic,ranjith2019field}.

\subsection{Inelastic Neutron Scattering}\label{INS}

\begin{figure}[tb]
\includegraphics[width=0.8\linewidth]{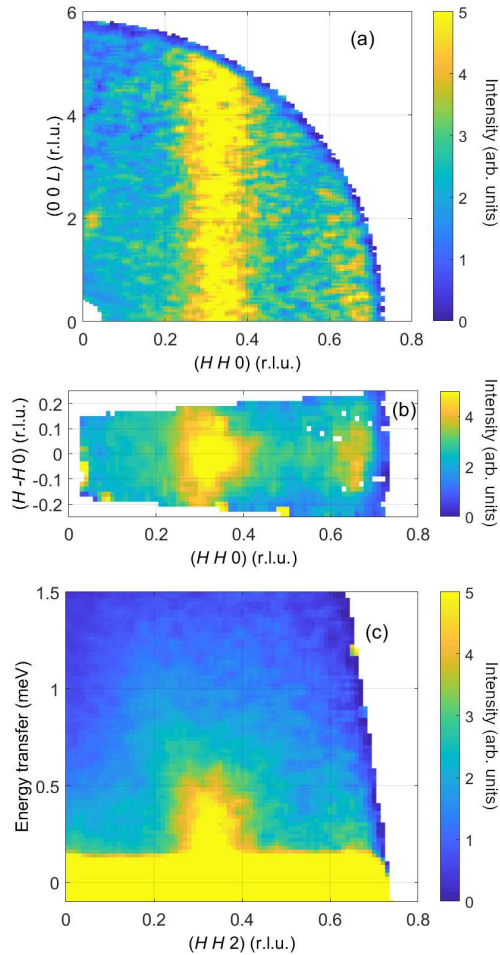}
\caption {(a,b) Constant energy plots in ($a^{\ast}c^{\ast}$) plane (a) and ($a^{\ast}b^{\ast}$) plane (b) at $T=2$~K. The scattering intensity was integrated within $\Delta E = [0.2, 0.4]$~meV. The data in (b) were symmetrized along the ($H$~-$H$~0) direction.  Axis values are in reciprocal lattice units (r.l.u.). (c) Low-energy excitation spectrum of CsYbSe$_2$ at $T=2$~K. Energy slice is taken along the $(H H 2)$ direction with ($H$~-$H$~0) and (0~0~$L$) integrated over the range [-0.1, 0.1] and [1,4] r.l.u., respectively.}
\label{ins_fig}
\end{figure}

The eight-fold degenerate $J = 7/2$ ($L = 3$, $S = 1/2$) multiplet ($2J + 1 = 8$) of Yb$^{3+}$ is split into four Kramers doublet states.
First, we probed for possible low-energy CEF excitations of the Yb$^{3+}$ ions using neutron incident energy $E_i=25$~meV.
We did not find any excited CEF levels in the energy range below 20~meV ($\sim$230~K), which is in agreement with the results obtained for other Yb-based triangle-lattice compounds, YbMgGaO$_4$ ($E_1=38$~meV)~\cite{paddison2017continuous} and NaYbS$_2$ ($E_1=23$~meV)~\cite{baenitz2018naybs}.
This is also in agreement with our heat capacity measurements where no Schottky anomaly is seen.
Since the ground state is well separated from the other excited CEF levels, an effective spin-1/2 description is indeed appropriate at low temperature and magnetic properties are dominated by the ground-state Kramers doublet.

The experimental INS spectra for CsYbSe$_2$ are presented in Fig.~\ref{ins_fig} as false color plots of the neutron intensity without any background subtraction.
Constant-energy slice in the ($a^{\ast}c^{\ast}$) plane, taken around energy $E=0.3$~meV [Fig.~\ref{ins_fig}(a)], shows clear intensity modulation along the ($HH$) direction.
For the magnetic planes decoupled along $c$, the dispersion of the excitation would be expected to be independent of $L$.
The data indeed indicate the absence of dispersion along $L$, implying that the inter-plane coupling is much weaker than the intra-plane correlations, in full agreement with the magnetization measurements.

The out-of-plane detector coverage of the CNCS TOF spectrometer is $\pm{15}^{\circ}$, so that a limited $Q$-range in the ($a^{\ast}b^{\ast}$) plane could also be accessed.
Fig.~\ref{ins_fig}(b) depicts the momentum dependence of an inelastic magnetic scattering for energy range from 0.2 to 0.4~meV.
Note that we symmetrized data along the ($H$~-$H$~0) direction in order to get large $Q$-coverage.
The strongly dispersive scattering centered at the (1/3~1/3~0) point is clearly seen, suggesting a 120$^{\circ}$ non-collinear 2D-like spin correlations.

Fig.~\ref{ins_fig}(c) illustrates the energy dependence of the scattering intensity along ($H~H$) direction, which reveals broad excitations originated from the (1/3~1/3~$L$).
The elastic scattering exhibits no magnetic Bragg peaks, confirming the absence of a long-range magnetic order.
The excitations are gapless within the energy resolution and have a bandwidth of about 0.6~meV.
In the highly degenerate spin-liquid ground state of our material, the magnetic excitations are over-damped away from the center of magnetic Brillouin zone, similar to the case of quantum spin-ice compound Yb$_2$Ti$_2$O$_7$~\cite{Ross}.
We note that the magnetic excitations in the CsYbSe$_2$ are in strong contrast to the data obtained for the other triangular-lattice compound Ba$_3$CoSb$_2$O$_9$, where well formed dispersion branches of single-magnon excitations and dispersive continua have been observed~\cite{Ito2017}.

\section{Conclusion}
Using the experimental results of magnetization, heat capacity and inelastic neutron scattering on CsYbSe$_2$ single crystals, we show that there is significant evidence to suggest that CsYbSe$_2$ provides a natural realization of the quantum spin liquid model at low magnetic fields.
The magnetization and heat capacity show the absence of conventional magnetic order and spin freezing down to $T=0.4$~K at zero field.
A magnetization plateau at one-third of the saturation magnetization is found at temperatures below 1.2~K, where temperature dependence of heat capacity confirms the magnetic field induced quantum phase transition.
The central features of the observed inelastic magnetic scattering, that is the low-energy gapless over-damped 2D excitation centered at the (1/3~1/3~0) point, are the essential experimental hallmark of the $S$=1/2 triangular-lattice QSL.
Our work also calls for further studies of the effects of magnetic field on the ground state and excitations, including mapping out dynamics in the 1/3-plateau.
This requires larger single-crystal samples and INS measurements in field at ultra low temperatures.
This work is in progress.
\newline
\newline

\section*{Acknowledgments}

We would like to thank D.~Pajerowski and S.~Nikitin for useful discussions.
The research at the Oak Ridge National Laboratory (ORNL) is supported by the U.S. Department of Energy (DOE), Office of Science, Basic Energy Sciences (BES), Materials Science and Engineering Division.
This research used resources at Spallation Neutron Source, a DOE Office of Science User Facility operated by ORNL.
X-ray Laue alignment was conducted at the Center for Nanophase Materials Sciences (CNMS) (CNMS2019-R18) at ORNL, which is a DOE Office of Science User Facility.
Work at Florida by J. S. Kim and G. R. S. supported by the US Department of Energy, Basic Energy Sciences, contract no. DE-FG02-86ER45268

%


\end{document}